\documentclass[slac_one]{revtex4}
\usepackage{graphicx}
\usepackage{fancyhdr}
\newcommand{\qsq}{\ensuremath{Q^2} }
\newcommand{\gevsq}{\ensuremath{\mathrm{GeV}^2} }

\newcommand{\xgobs}{\ensuremath{x_{\gamma}^{obs} }}
\newcommand{\gev}{{\rm\,GeV}}
\newcommand{\mev}{{\rm\,MeV}}

\begin{document}

\title{Strangeness Production and Hadron Spectroscopy at HERA}
%% Paper title goes here

\author{Christoph Grab\, \,  (for the H1 and ZEUS Collaborations)  }
\affiliation{
Institute for Particle Physics, ETH Zurich \\
8093 Zurich, Switzerland  }
\vspace{-6mm}
\begin{abstract}
An overview of the
recent results on strangeness production and spectroscopy from the electron-proton
collider experiments H1 and ZEUS at HERA is presented.
Production of particles with light and strange 
quarks is discussed and compared with both theoretical predictions and with
data from $e^+e^-$ experiments.
Measurements in the charm sector cover 
studies of the radially and orbitally excited charm states.
Finally, the investigation of exotic states in the strangeness 
sector at HERA is reviewed.
\end{abstract}

\maketitle

\thispagestyle{fancy}


%%%%%%%%%%%%%%%%%%%%%%%%%%%%%%%%%%%%%%%%%%%%%%%%%%%%%%%%%%%%%%%%%%
\section{INTRODUCTION } % Section title should be in all capitals.
\vspace{-4mm}
The two experiments H1 and ZEUS measured 
collisions of electrons\footnote{
HERA was operated with both electron and positron beams.
A reference to electron implies a reference to either
electron or positron. 
Most of the results shown were obtained in $e^+p$ collisions.
Also, charge conjugate states are always implicitly included.
}
with  protons at the HERA  collider
from 1992 to June 2007.
The main focus of interest at HERA lies in the study of the structure of the
proton, in detailed investigations of the various aspects of perturbative
QCD and in searches for new phenomena.
Relevant in the context of this conference are the detailed studies of
production processes and the differences therein between
e.g. mesons, baryons and antiparticles. 
In addition, details about the non-perturbative aspects of fragmentation
are investigated.
Search and studies of the production of excited and more ``exotic'' states
in the $ep$ environment, such as pentaquarks or glueballs constitute
another part of the HERA physics portfolio.
Overall, the HERA $ep$ results complement the findings of other experiments and
help to build phenomenological models with strong predictive power,
applicable at other machines.

%%%%%%%%%%%%%%%%%%%%%%%%%%%%%%%%%%%%%%%%%%%%%%%%%%%%%%%%%%%%%%%%%%%%%%%%%%

%The kinematics of the $ep$ scattering reaction 
%are described elsewhere.
At HERA two different regimes in the photon momentum transfer squared 
\qsq\ are distinguished:
{\it ``photoproduction''} denotes low $Q^2 \approx 0$ \gevsq, 
%also termed $\gamma p$, where the photon emitted 
%from the electron is quasi-real; 
and {\it ``deep inelastic scattering (DIS)''} refers to high \qsq\ 
(in practice $Q^2  \geq 2 \gevsq$).
In the case of photoproduction, the additional variable \xgobs\ 
describes the momentum fraction of the photon, that participates in the
hard interaction. 
%It is used to experimentally distinguish
%direct photon ($\xgobs \approx 1$) and resolved photon
%($\xgobs < 1 $) processes.
The total integrated luminosity available for physics at HERA
amounts to about 1 fb$^{-1}$.
The results presented below are based on fractions of this.

%%%%%%%%%%%%%%%%%%%%%%%%%%%%%%%%%%%%%%%%%%%%%%%%%%%%%%%%%%%%%%%%%%%%%%%%%%

%%%%%%%%%%%%%%%%%%%%%%%%%%%%%%%%%%%%%%%%%%%%%%%%%%%%%%%%%%%%%%%%%%%%%%%%%%
%
\section{STRANGE PARTICLE PRODUCTION}
\vspace{-4mm}
%\subsection{Production Cross Sections of $K$-Mesons and $\Lambda$-Baryons}
{\bf Production Cross Sections of $K$-Mesons and $\Lambda$-Baryons}\\
The production of strange quarks can proceed 
perturbatively by the boson-gluon fusion process 
($\gamma g \rightarrow s \bar{s}$) and by gluon splitting in parton showers. 
In addition, the proton parton densities and the non-perturbative string fragmentation
can be sources of strange quarks. The strange hadrons 
are then produced in the hadronisation process, or through decays
of higher mass states.
%
%%%%%%%%%%%%%%%%%%%%%%%%%%%%%%%%%%%%%%%%%%%%%%%%%%%%%%%%%%%%%%%%%%%%%%%%%%
\begin{figure}[hbt]
    \begin{center}
          \includegraphics[width=56mm,height=50mm]{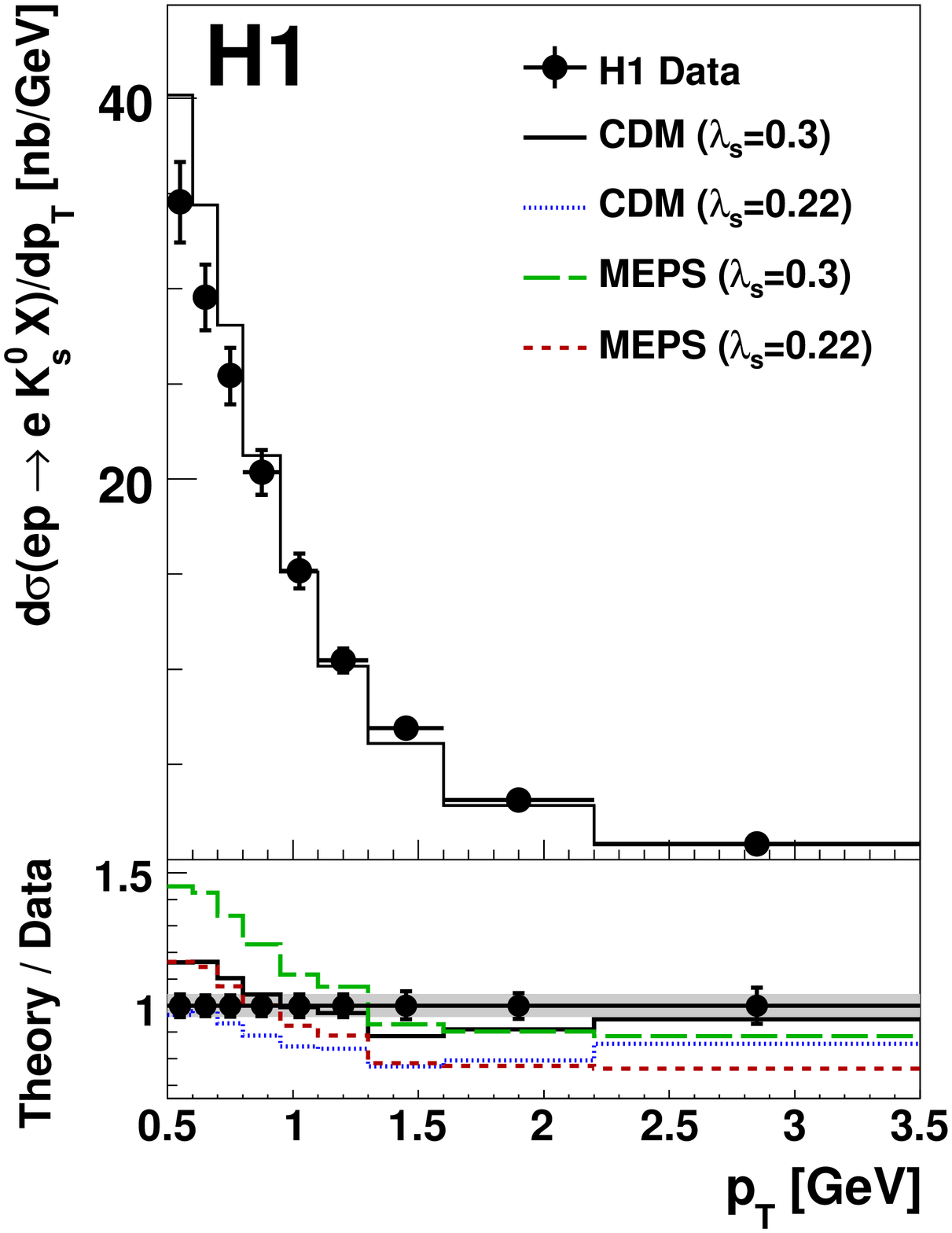}
          \includegraphics[width=56mm,height=50mm]{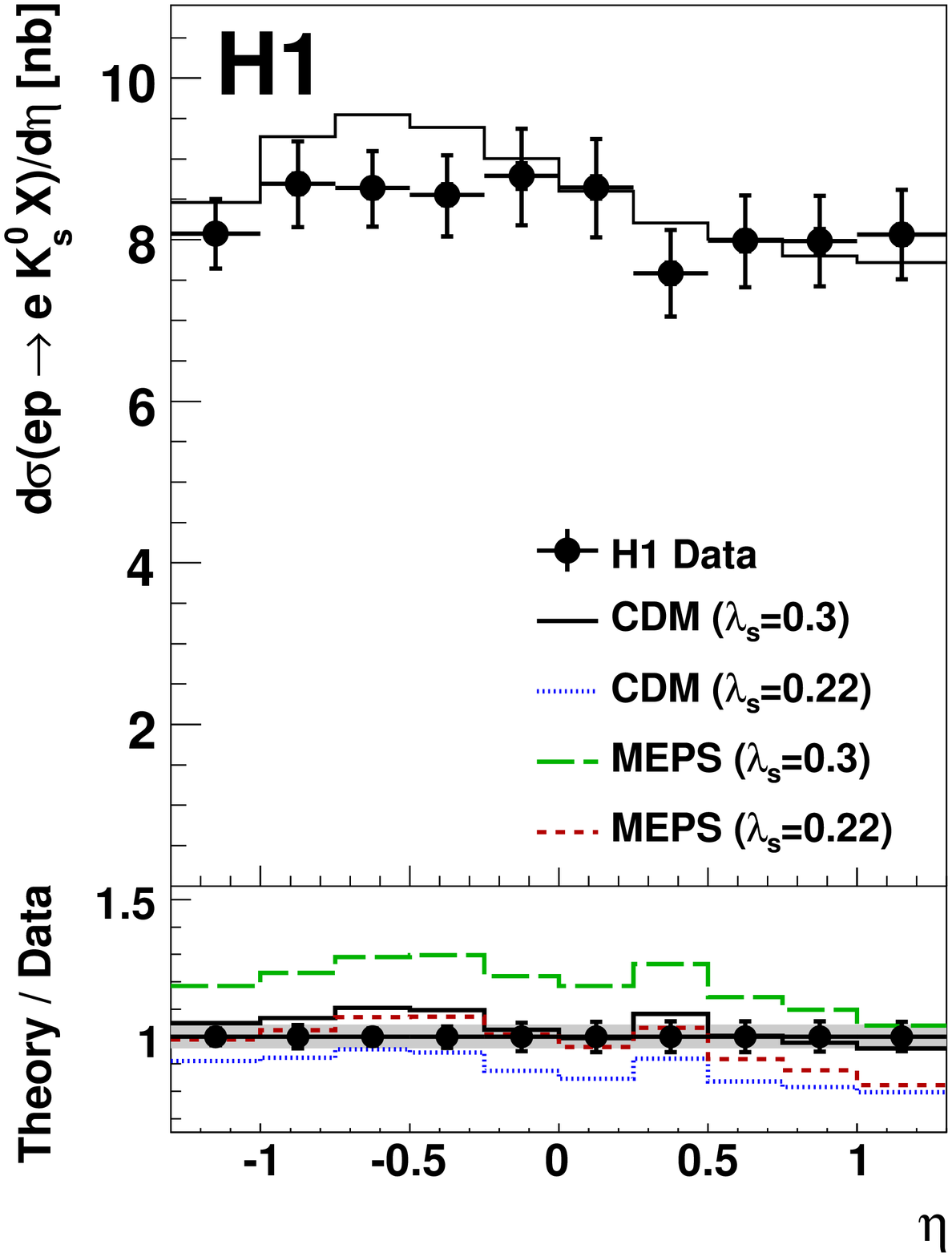}
          \includegraphics[width=56mm,height=50mm]{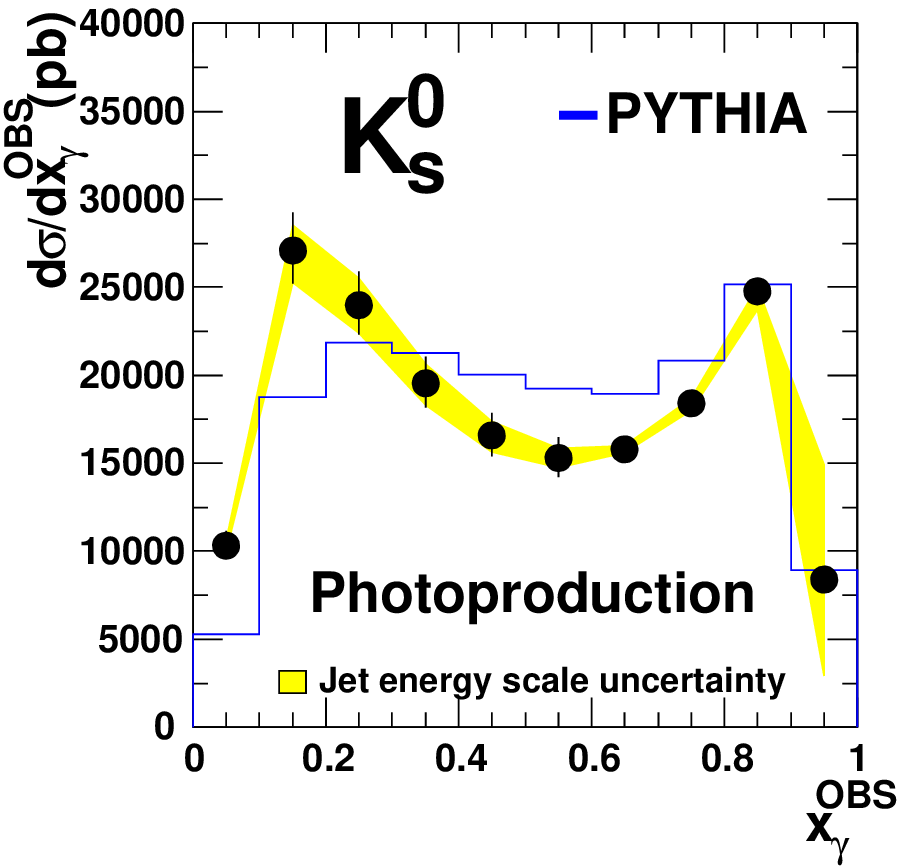}
    \end{center}\vspace{-6mm}
\caption{\it Distribution of the $K^{0}_{S}$ production cross 
section 
as a function of a) $P_{T}$, b) $\eta$ in the laboratory frame for 
the DIS  \qsq\ region by H1~\cite{H1-strange08},
and c) $x_{\gamma}^{obs}$ in $\gamma p$ by ZEUS~\cite{Chekanov:2006wz}.
%The band in c) indicates the jet energy scale uncertainty.
}
\label{fig:ks-spectra}
\end{figure}
%%%%%%%%%%%%%%%%%%%%%%%%%%%%%%%%%%%%%%%%%%%%%%%%%%%%%%%%%%%%%%%%%%%%%%%%%%
%
A study of $K_S$  and $\Lambda$ production 
was reported by the  H1 collaboration~\cite{H1-strange08} 
in DIS for $2 < Q^2 < 100 \gev^2$, and by the 
ZEUS collaboration\cite{Chekanov:2006wz}
in DIS with $Q^2 > 25 \gev^2$, with $5 < Q^2 < 25 \gev^2$,
and in photoproduction.
The various aspects of the production studied are:
single differential cross sections of $K_S$  and $\Lambda$,
baryon-antibaryon asymmetry, 
baryon-to-meson ratio, ratio of strange-to-light hadrons,
and the $\Lambda$ and $\bar{\Lambda}$ transverse spin 
polarization. 
The spectra include all sources, 
i.e. direct production and all resonance decays.
%Results
As an example the $K_S^0$ spectra are shown in Figure~\ref{fig:ks-spectra}
as function of $P_{T}$ and $\eta$ in DIS, and
of $x_{\gamma}^{obs}$ in $\gamma p$.

\begin{figure}[hbt]
    \begin{center}
          \includegraphics[width=56mm,height=56mm]{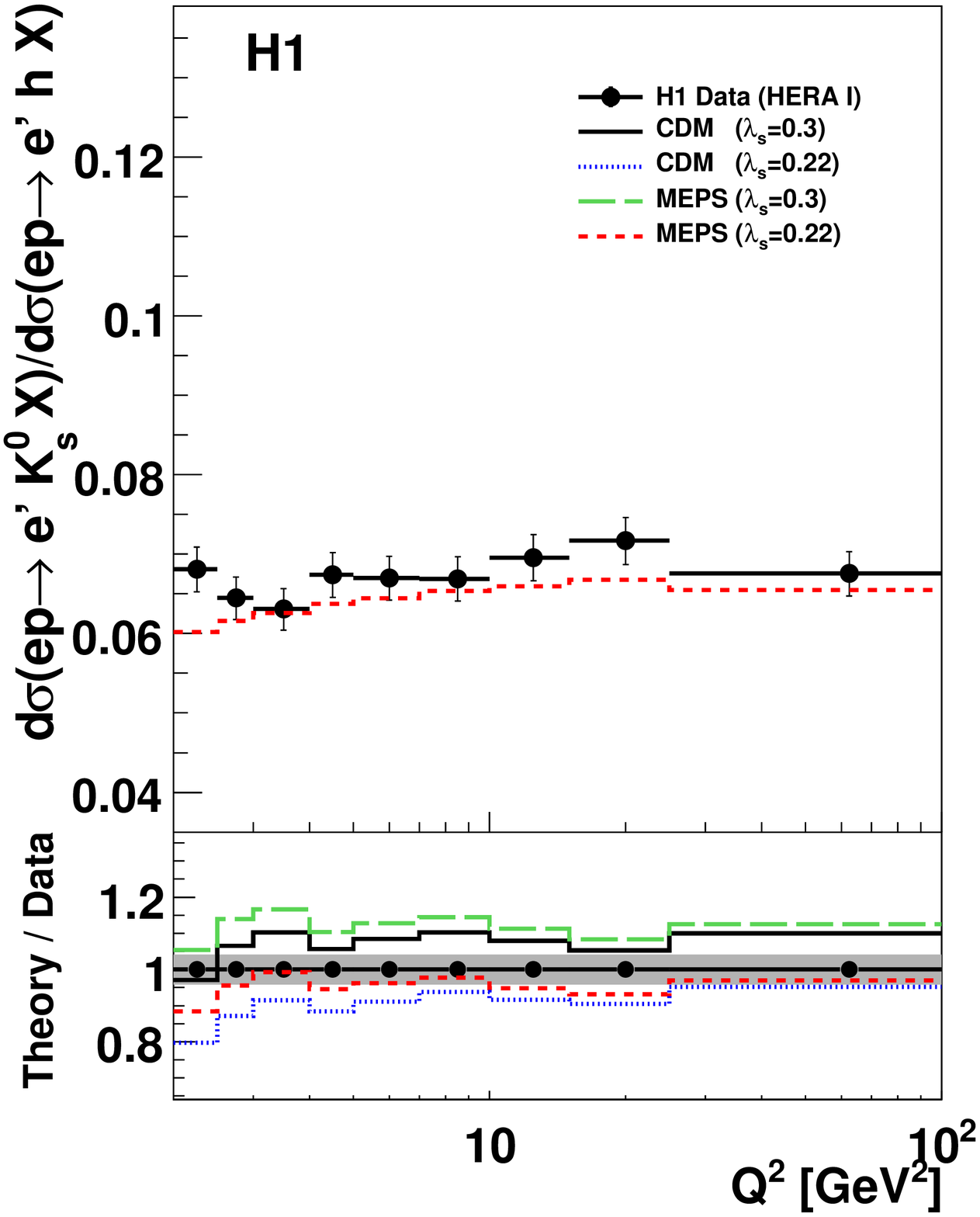}
          \includegraphics[width=56mm,height=56mm]{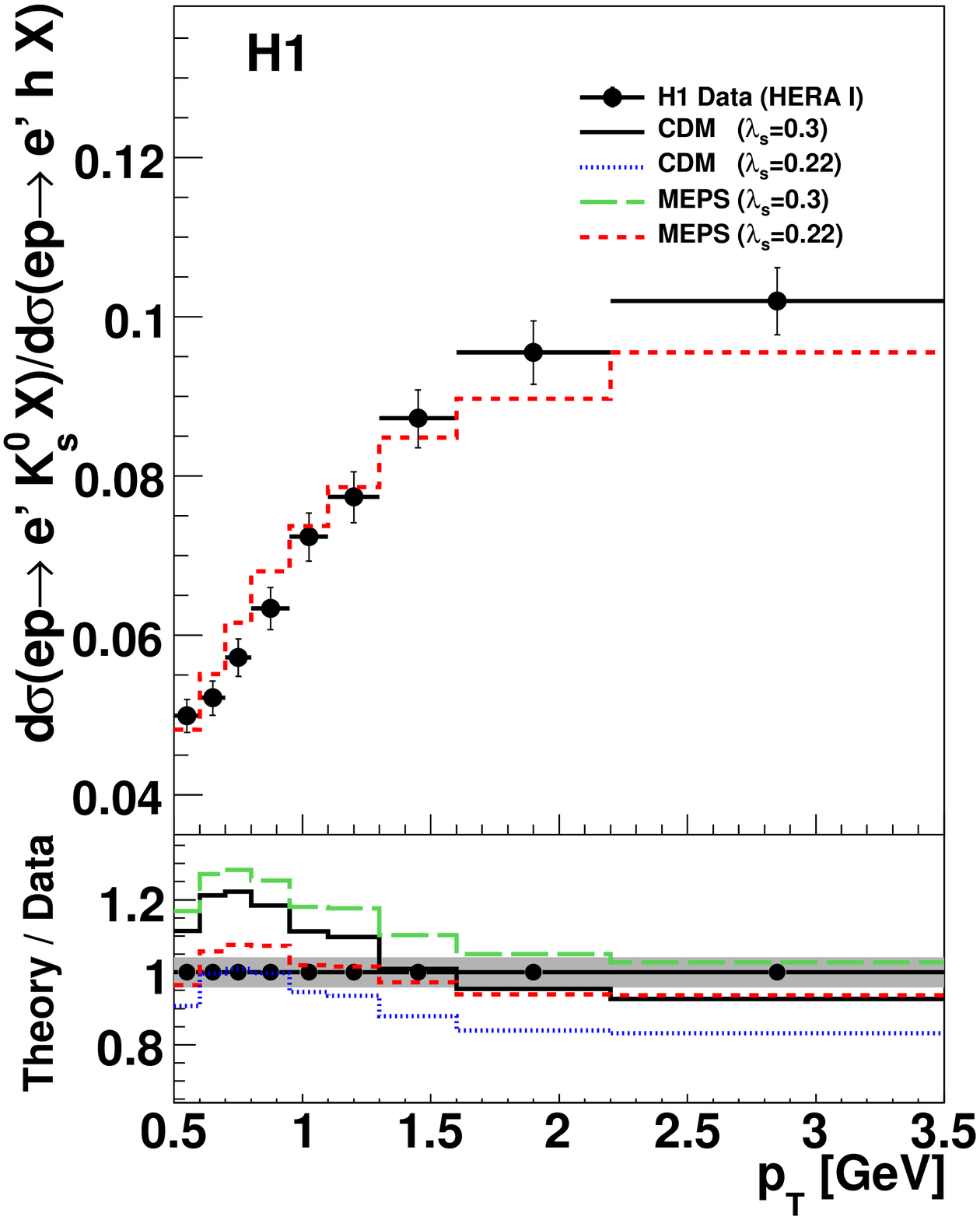}
          \includegraphics[width=56mm,height=56mm]{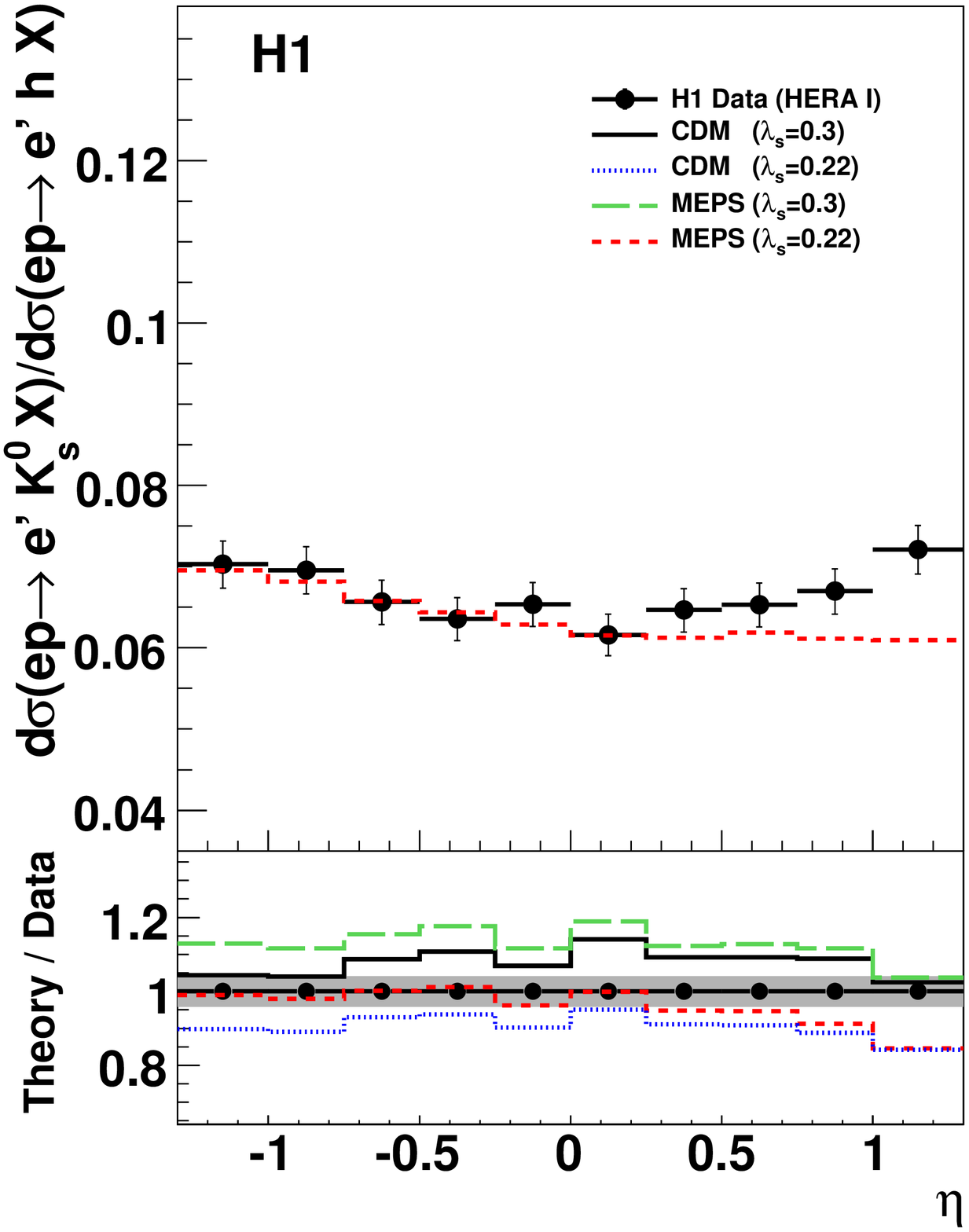}
    \end{center}\vspace{-6mm}
\caption{\it Distribution of the 
$K^{0}_{S}$ to charged particles ratio 
${N(K^{0}_{S})}/{N_\mathrm{ch}}$ as a function of 
a) \qsq\, b) $P_{T}$ and c) $\eta$ for
the DIS region (H1~\cite{H1-strange08}).
}
\label{fig:s-l-ratio}
\end{figure}

A comparison of the various measured single differential spectra 
with  predictions based on LO QCD models shows a good overall 
agreement, when using a strangeness suppression 
factor of $\lambda_S =0.3$.
However, a closer look reveals several differences between theory and data.
Similar general conclusions can be drawn from the baryon-to-meson ratio distributions.
%(see Figure~\ref{fig:b-m-ratio}).
%
For instance, the shape of the  $x_{\gamma}^{obs}$ distribution
in $\gamma p$ is not properly described by the PYTHIA simulation in both cases.
%For the direct contribution, the baryon-to-meson
%ratio in $\gamma p$ is similar to the one
%in DIS and also similar to $e^+e^-$ data.
The general features of the strange-to-light hadron ratio 
measurements (see Figure~\ref{fig:s-l-ratio}) are 
pretty well described by the simulations, provided
a lower $\lambda_S$ value such as $0.22$ is chosen.
There was no asymmetry observed between $\Lambda$ and ${\bar{\Lambda}}$,
which indicates a similar production process for baryons and antibaryons.
Studies of angular distributions in the $\Lambda$ decays did not reveal 
any non-zero transverse polarization in the $\Lambda$ or ${\bar{\Lambda}}$ production.
%
%%%%%%%%%%%%%%%%%%%%%%%%%%%%%%%%%%%%%%%%%%%%%%%%%%%%%%%%%%%%%%%%%%%%%%%%%%
%see Figure~\ref{fig:s-l-ratio})
%%%%%%%Kstar
A first measurement of the production cross section 
of the excited $K^*(892)$
kaon state, detected through the decay
$K^*(892) \to K_S^0 \pi$ in the DIS region $5 < \qsq < 100\ \gevsq$ and $0.1 < y < 0.6$
was presented by the H1 collaboration~\cite{H1-kstar08}.
%As an example, Figure~\ref{fig:kstar} shows the observed signal distribution
%and the $K^*(892)$ production cross sections in the lab frame.
%
Comparisons of the results with predictions of leading order Monte Carlo 
models matched to parton showers exhibit a picture, which is well 
consistent with the conclusions drawn from the  $K_S, \Lambda$ production 
measurements.
%
%%%%%%%%%%%%%%%%%%%%%%%%%%%%%%%%%%%%%%%%%%%%%%%%%%%%%%%%%%%%%%%%%%%%%%%%%%
%\begin{figure}[hbt]
%    \begin{center}
%          \includegraphics[width=56mm,height=50mm]{figs/InvMass_KStarAll.eps}
%          \includegraphics[width=56mm,height=50mm]{figs/XSection_Pt.eps}
%          \includegraphics[width=56mm,height=50mm]{figs/XSection_Eta.eps}
%    \end{center}\vspace{-6mm}
%\caption{\it a) Invariant mass distribution of $K_S^0 \pi$ from the
%decay $K^*(892) \to K_S^0 \pi$, and the $K_S^0 \pi$ 
%production cross sections in the laboratory frame
%as a function of b) $P_{T}$ and c) $\eta$
%(H1~\cite{H1-kstar08}).
%}
%\label{fig:kstar}
%\end{figure}
%%%%%%%%%%%%%%%%%%%%%%%%%%%%%%%%%%%%%%%%%%%%%%%%%%%%%%%%%%%%%%%%%%%%%%%%%%
% K,L summary:%%%%%%%%%%%%%%%%%%%%%%%%%%%%%%%%%%%%%%%%
%
In summary,
% the H1 and ZEUS measurements yield quite consistent conclusions.
%It is found, 
%that the overall features of the $K_S, \Lambda$ production and their ratios 
%are quite reasonably described by the simulation.
a single
strangeness suppression factor $\lambda_S$ is not sufficient for a consistent
description of all distributions in detail.
This is partly due to the fact, that in various regions of phase
space,  the relative contributions of direct production of strange to other quarks 
differ; in particular at larger scales (e.g. \qsq, $p_T$) the charm contributions
become substantial and need to be properly included into the description.\\

%%%%%%%%%%%%%%%%%%%%%%%%%%%%%%%%%%%%%%%%%%%%%%%%%%%%%%%%%%%%%%%%%%%%%%%%%%
%\subsection{Bose Einstein Correlations of Charged and Neutral Kaons}

\begin{figure}[hbt]
    \begin{center}
          \includegraphics[width=50mm,height=40mm]{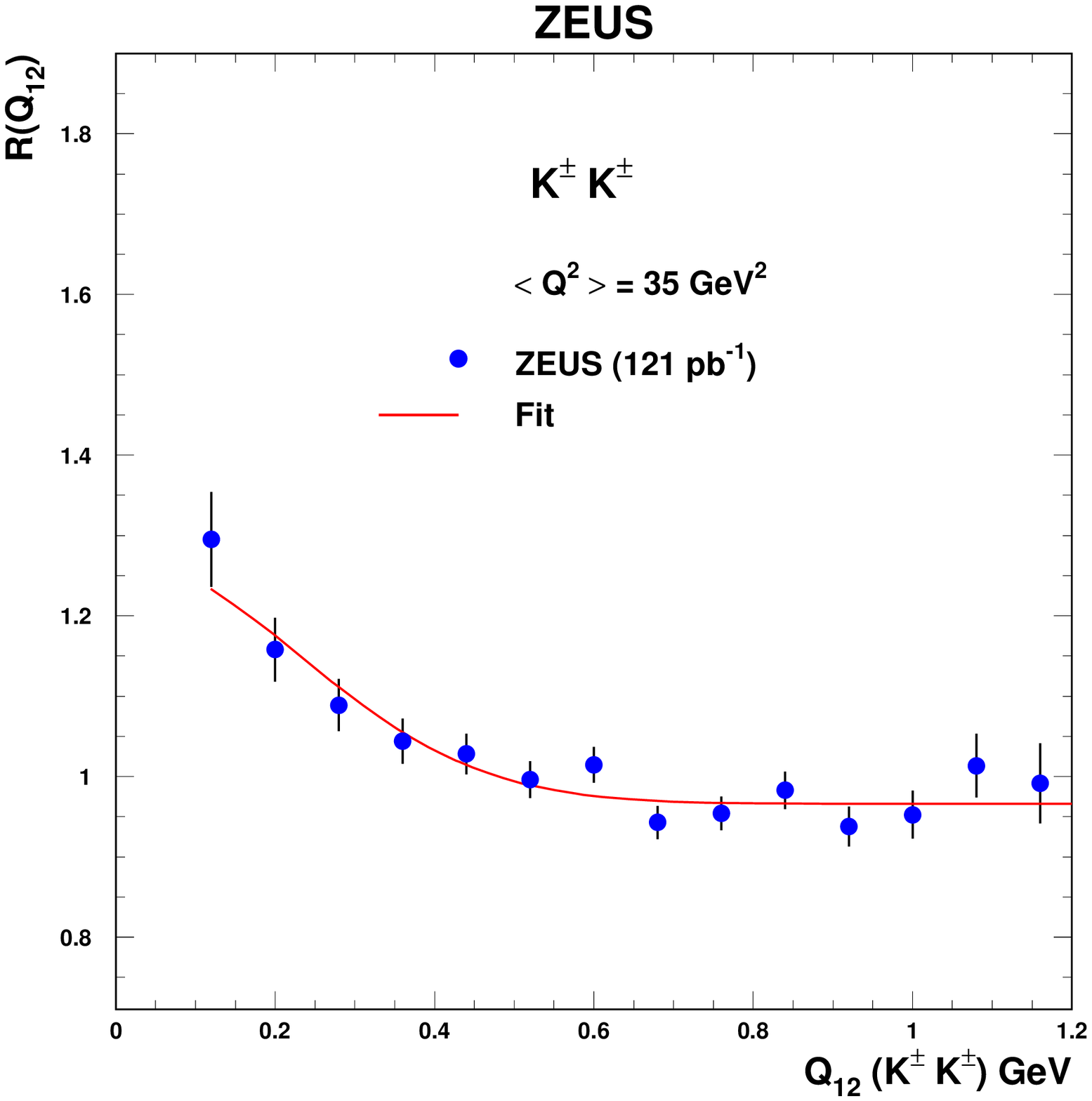}
          \includegraphics[width=50mm,height=40mm]{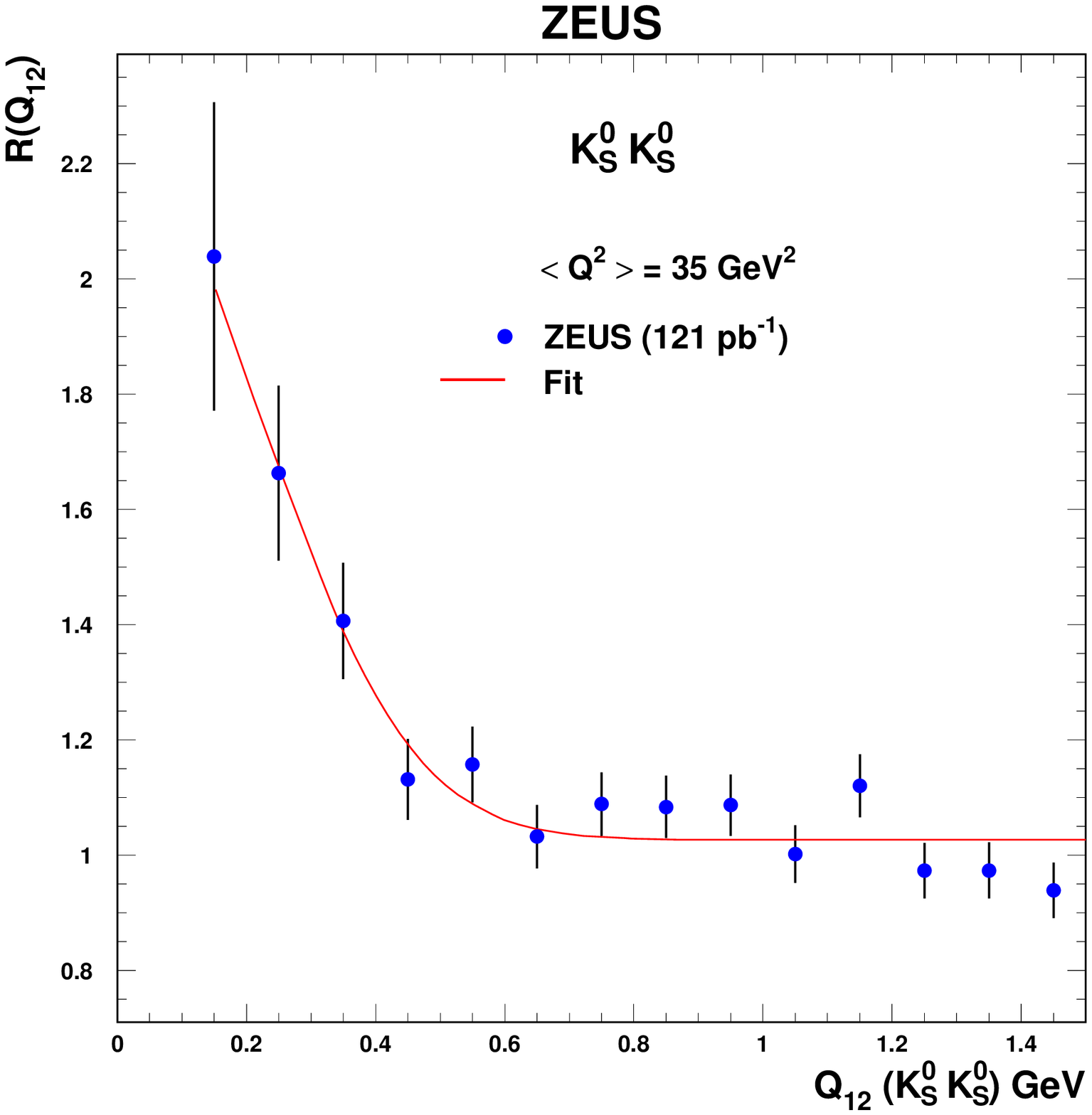}
          \includegraphics[width=50mm,height=36mm]{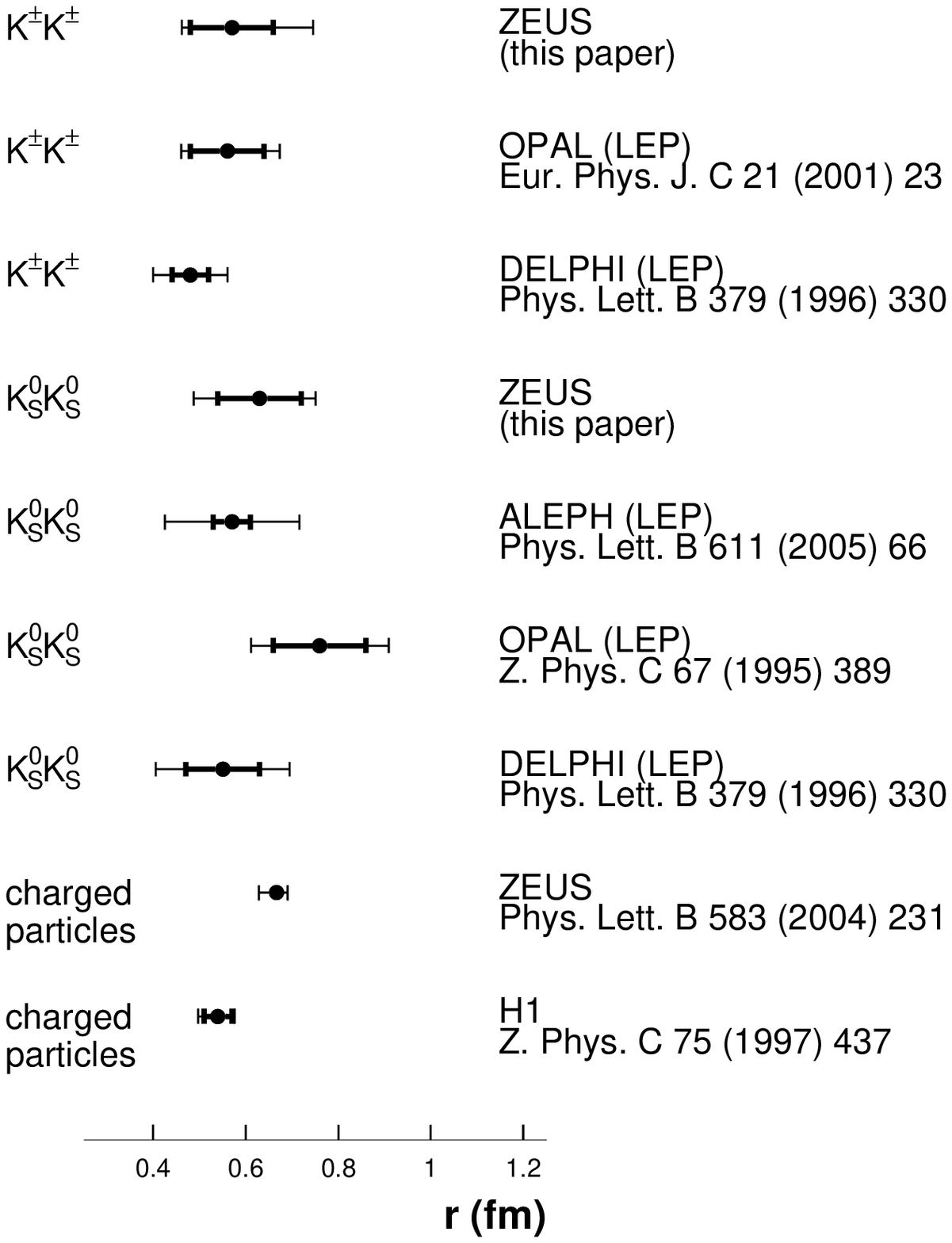}
    \end{center}\vspace{-6mm}
\caption{\it
The two-particle correlation function as measured by ZEUS~\cite{chekanov:2007ev}
for a) charged kaons  and b) neutral kaons.
In c) the measured source radii are compared to published values from 
LEP and HERA. }
\label{fig:zeus-bec}
\end{figure}

{\bf Bose Einstein Correlations of Charged and Neutral Kaons} \\
Bose-Einstein correlations (BEC) have been used in particle physics as 
a method of determining the size and the shape of a particle emitting source.
BEC originate from the symmetrization of the 
two-particle wave function of identical bosons and lead to an enhancement 
of boson pairs emitted with small relative momenta.
The ZEUS collaboration reported~\cite{chekanov:2007ev}
 on studies of BEC for pairs of identical 
kaons, both neutral and charged, in DIS.
ZEUS studied the two-particle correlation function $R(Q_{12})$
as a function of              
the four-momentum difference of the kaon pairs,                                 
$Q_{12} = \sqrt{-(\textrm{p}_1-\textrm{p}_2)^2}$,                               
assuming a Gaussian shape for the particle source.                             
The actual correlation function  $R(Q_{12})$ was calculated by means of a double
ratio, comparing with a reference sample free from the BOSE-Einstein effect.
The measured distributions are shown in Figure~\ref{fig:zeus-bec}.

The values of the radius of the production volume, {\it r}, and of the          
correlation strength, $\lambda$, were determined from these distributions.
While the charged kaon data sample yields quite precise results, 
the neutral kaon data suffer from larger systematic errors due
to background contributions from the $f_0(980)\rightarrow K^0_SK^0_S$ decays.

The  radii for charged and neutral kaons are found to be in agreement within
errors, similar to the ones measured in DIS and   
are also seen to be consistent with those obtained at LEP. 
A comparison of these results with other published values is shown 
in Figure~\ref{fig:zeus-bec}c.
%
%%%%%%%%%%%%%%%%%%%%%%%%%%%%%%%%%%%%%%%%%%%%%%%%%%%%%%%%%%%%%%%%%%%%%%%%%%
%
%
%%%%%%%%%%%%%%%%%%%%%%%%%%%%%%%%%%%%%%%%%%%%%%%%%%%%%%%%%%%%%%%%%%%%%%%%%%
\vspace{-8mm}
%
%{\bf \large CHARMED PARTICLE PRODUCTION} \\
\section{CHARMED PARTICLE PRODUCTION}
\vspace{-4mm}
%
%The theoretical description of inelastic charm quark production at HERA
%is based on perturbative QCD (pQCD).
%In leading order (LO) the direct process of 
%{\em photon-gluon fusion} is the dominant contribution,
%i.e.\ $\gamma g \to c \bar{c}$.
%In photoproduction, resolved photon interactions contribute as well, 
%i.e.  $g g  \to c \bar{c} , q \bar{q}  \to c \bar{c}$, 
%and are described with the help of a photon structure function.
%Beyond LO {\it only the  sum}\ of 
%direct and resolved processes is a well-defined quantity.
%The charm quarks from the hard interaction hadronise either in 
%``open states'' (e.g. $D$-mesons
%$\Lambda_c$ ...) or in ``hidden ${\bar c}c$ states'', 
%such as $J/ \Psi$.
%The hadronisation is then described by non-perturbative models, that
%contain various parameters which need to be determined by experiments.
%
%%%%%%%%%%%%%%%%%%%%%%%%%%%%%%%%%%%%%%%%%%%%%%%%%%%%%%%%%%%%%%%%%%%%%%%%%%
% \subsection{Excited States}
The ZEUS collaboration has reported\cite{Zeus-excite08}
the observation
of excited $D$ and $D_s$ mesons in the decay chains
$D_1(2420)^0, \ D^*_2(2460)^0 \to D^{*\pm} \pi^{\mp},  D^{\pm} \pi^{\mp}$  
and $D^{\pm}_{s1}(2536) \to D^{*\pm} K^0_S,\  D^0 K^{\pm}$.
Based on a sample of $\approx 57000\ D^{*}$ and $\approx 20400\ D^{\pm}$ 
and $\approx 22000\ D^{0}$ mesons, the decays are studied in detail.
%The resulting observed mass difference spectra are shown in 
%Figure~\ref{fig:charm-excite}.
From a total of 3110 $D_1(2420)^0$, $1560\ D^*_2(2460)^0$ and 
 236 $D^{\pm}_{s1}(2536)$ states, properties like mass,
widths and decay fractions are determined.
In general, the measured properties agree  well with the
PDG values~\cite{PDG}, 
except the width of the $D_1(2420)^0$ state, which
is found to be larger than the world average.
The measured fragmentation fractions for these states $f_c$ 
in $ep$ are 
%indicated in Table I. They are 
found to be
compatible with the ones observed in $e^+e^-$,
supporting the universality of fragmentation.

In addition, an helicity analysis of the angular distributions
between the $K_S$ and the $\pi_s$ in the $D^*$ restframe for the
$D^{\pm}_{s1}(2536)$ yielded a helicity parameter 
of $h(D^{\pm}_{s1}) = -0.74 ^{+0.23}_{-0.17} (stat) ^{+0.06}_{-0.05} (syst)$.
This non-zero value of the helicity parameter $h$ suggests 
the state to be rather a mixture
of two $1^+$ states, e.g. to contain contributions from both $S$-
and $D$-waves, as $h=0 (3)$ is expected for a pure $S (D)$-state.

A search for the radially excited state  $D^{*'+}(2640)$ does not
show a significant signal, and thus leads to the presently best
upper limit on the charm branching fraction $f(c \to D^{*'+}(2640))$.
%(see Table I).
%%
%\begin{table}[t]
%\begin{center}
%\begin{tabular}{|l|l|}
%\hline
%  $f(c \to D^0_1(2420))$ = & $3.5 \pm 0.4 ^{+0.4}_{-0.6} \% $   \\
%\hline
%  $f(c \to D^{*0}_2(2460))$ = & $3.8 \pm 0.7 ^{+0.5}_{-0.6} \% $    \\
%\hline
% $f(c \to D^+_{s1}(2536) $) = & $1.11 \pm 0.16 ^{+0.08}_{-0.10}  \% $    \\
%\hline
% $f(c \to D^{*'+}(2640))\cdot BR $ & $< 0.4 $ at $95 \% $ c.l.     \\
%\hline 
%\end{tabular}
%\label{tab:zeus-tabx}
%\caption{Fragmentation fractions  $f_c$ of excited $D,D_s$ mesons
%in \%, measured by ZEUS\cite{Zeus-excite08}. 
%}
%\end{center}
%\end{table}
%%
%\begin{figure}[hbt]
%    \begin{center}
%          \includegraphics[height=50mm, width=80mm]{figs/DESY-08-093_3.eps}
%          \includegraphics[height=50mm width=80mm]{figs/DESY-08-093_7.eps}
%    \end{center}
%\vspace{-6mm}
%\caption{\it Mass difference distributions\cite{Zeus-excite08} for 
%a) charmed excited P-wave meson decays
% $D_1(2420)^0, \ D^*_2(2460)^0 \to D^{*\pm} \pi^{\mp}$ (upper) and
%$\to D^{\pm} \pi^{\mp}$ (lower);
%and b) the charm-strange particle decays
%$D^{\pm}_{s1}(2536) \to D^{*\pm} K^0_S$ (upper) and 
%$\to D^0 K^{\pm}$ (lower).
%}
%\label{fig:charm-excite}
%\end{figure}
%

%%%%%%%%%%%%%%%%%%%%%%%%%%%%%%%%%%%%%%%%%%%%%%%%%%%%%%%%%%%%%%%%%%%%%%%%%%
%
\section{EXOTIC STATES IN THE STRANGENESS SECTOR}
\vspace{-4mm}
%
%\subsection{Strange Pentaquark States}
{\bf Strange Pentaquark States} \\
Both experiments, H1 and ZEUS, have searched for the
doubly strange pentaquark candidate states 
$\Xi^{++/00}$ in both charge combinations (doubly charged or
neutral) decaying in 
$\Xi^{++/00} \to \Xi^+ \pi^+$ and $\to \Xi^{\pm} \pi^{\mp}$.
Such states could be interpreted as the
$\Xi^{--}_{5q}\,$ ($S$\,=\,$-2$, $I_3=-3/2$) and
the $\Xi^{0}_{5q}\,$ ($S$\,=\,$-2$, $I_3=+1/2$)
members of the isospin $3/2$ quartet $\Xi_{3/2}$ in the 
anti-decuplet~\cite{pq-theory}.
The invariant $\Xi \pi$ mass spectra measured by the
H1\cite{Aktas:2007dd} and ZEUS\cite{Chekanov:2005at} collaborations
do not show any indication of a signal, apart from the
well known $\Xi(1530)^{0}\,$ baryon resonance.
%as shown in Figure~\ref{pentamm_limits}.
%
Therefore the non-observation of any resonance state 
in the mass range $1600 - 2300\ \mev$ in neither of the
charge combinations limits the production rate of a hypothetical 
$\Xi^{++/00}$ pentaquark to values of order $20\%$ on average 
at the $95\%$ C.L. relative to the well known $\Xi(1530)^{0}\,$ 
baryon production rate,
depending on the ($\Xi \pi$)-mass.\\

\begin{figure}
\begin{center}
    \includegraphics[width=40mm,height=50mm]{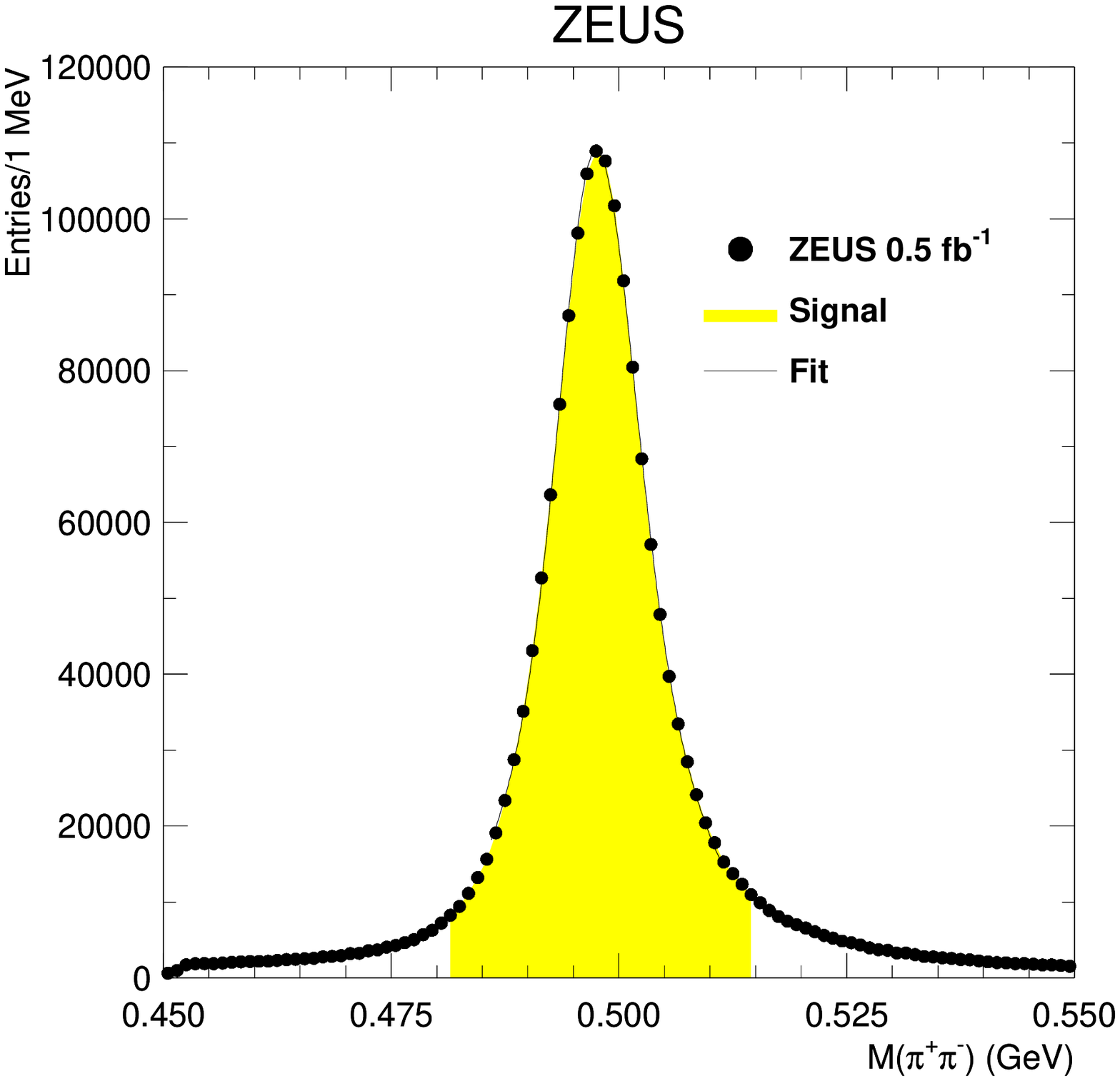}
    \includegraphics[width=80mm,height=50mm]{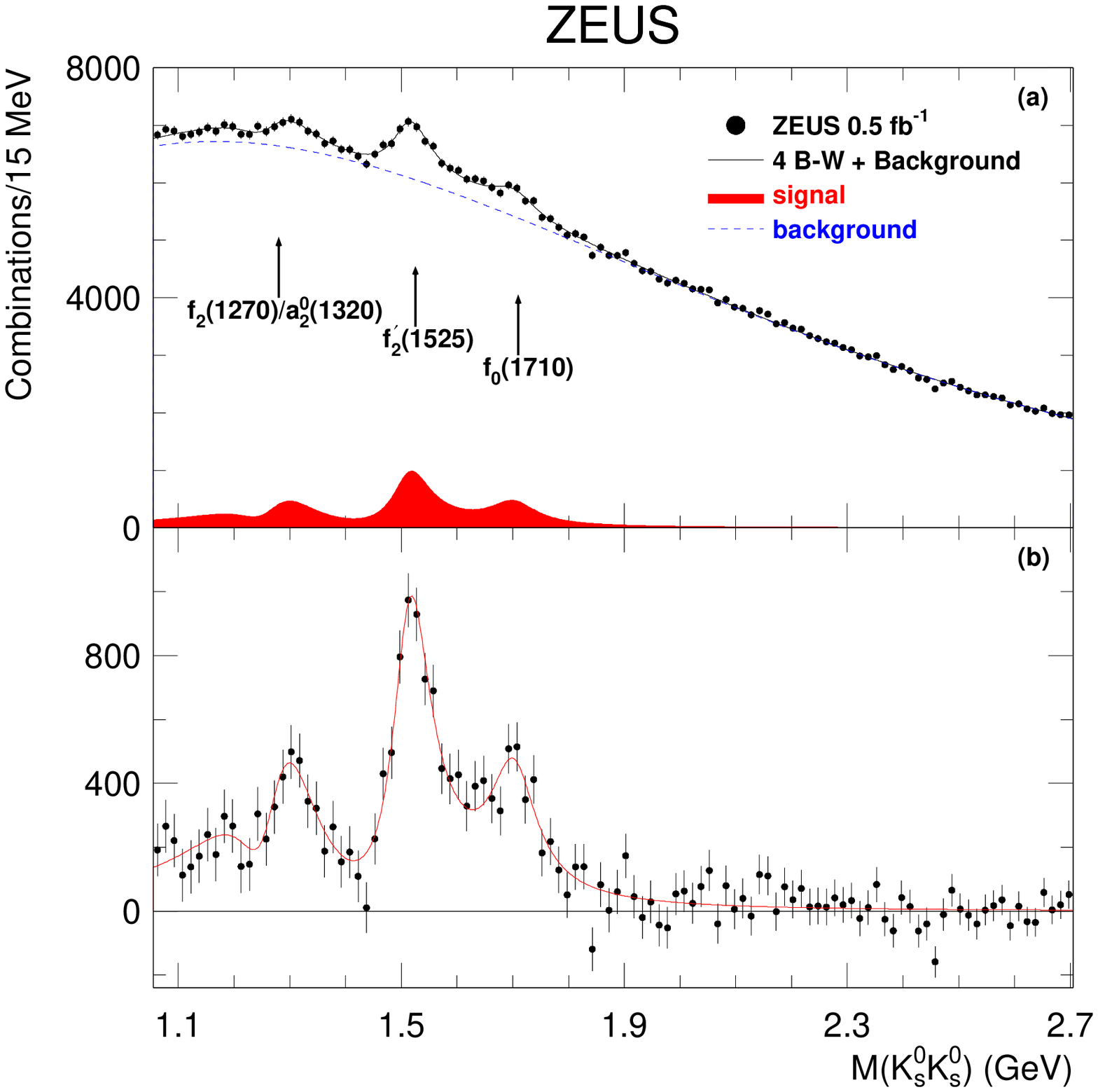}
\caption{The  invariant mass spectrum for a) $K^{0}_{S} \to \pi^+ \pi^-$ and
b) for $K^{0}_{S}K^{0}_{S}$ combinations in the
final states (ZEUS~\cite{Chekanov:ksks}).
}
\label{ksks-glueball}
\end{center}
\end{figure}

%\begin{figure}[htb]
%\begin{center}
%    \includegraphics[width=56mm,height=64mm]{figs/xi-Figure3.eps}
%    \includegraphics[width=56mm,height=64mm]{figs/xi-Figure4.eps}
%    \includegraphics[width=56mm,height=60mm]{figs/DESY-05-018_3.eps}
%\caption{The  invariant mass spectrum for
%a) the H1\cite{Aktas:2007dd} doubly charged combinations 
%$\Xi^- \pi^-$ and $\bar\Xi^+ \pi^+$ (upper part),
%b) the H1 neutral combinations $\Xi,^- \pi^+$ and $\bar\Xi^+ \pi^-$ (upper part),
%and c) the ZEUS\cite{Chekanov:2005at} sum of all four charge combinations.
%clearly visible is the well-known $\Xi(1530)^{0}\,$  baryon.
%The lower part shows the $95\%$ C.L. upper limit on the production ratio $R(M)$ 
%relative to the  $\Xi(1530)^{0}\,$  baryon,
%as a function of the mass $M$.
%}
%\label{pentamm_limits}
%\end{center}
%\end{figure}
%
%%%%%%%%%%%%%%%%%%%%%%%%%%%%%%%%%%%%%%%%%%%%%%%%%%
%\subsection{Glueball Candidates in the $K^{0}_{S}K^{0}_{S}$ Final State}
{\bf Glueball Candidates in the $K^{0}_{S}K^{0}_{S}$ Final State} \\
The ZEUS collaboration performed a detailed analysis of
the $K^{0}_{S}K^{0}_{S}$ final state~\cite{Chekanov:ksks}, 
based on a total sample of about 1.2 million reconstructed $K^{0}_{S}$
(see Figure~\ref{ksks-glueball}a).
The measured $K^{0}_{S}K^{0}_{S}$ invariant mass distribution is 
shown in Figure~\ref{ksks-glueball}b) and clearly exhibits three 
prominent structures.
The structures are elaborately analysed, employing multiple
relativistic Breit-Wigner functions and  allowing for interferences between
the various components. The background subtracted 
result of the fit is depicted in the lower part of Figure~\ref{ksks-glueball}b).
Clear evidence for the $f_2^{'}(1525)$ and the $f_2(1710)$ are observed
at the 5 sigma level. The masses and the widths of the different
contributions are measured. The widths are found to be consistent
with the PDG values~\cite{PDG}, whereas the masses of the $f_2^{'}(1525)$ 
and the $f_2(1710)$
are found to be slightly below the PDG values~\cite{PDG}.
The  $f_2(1710)$ state is found to be consistent with 
the lowest lying $J^{PC} = 0^{++}$ glueball candidate.
%
%%%%%%%%%%%%%%%%%%%%%%%%%%%%%%%%%%%%%%%%%%%%%%%%%%%%%%%%%%%%%%%%%%%%%%%%%%
%
\section{SUMMARY}
\vspace{-4mm}
The overall features of the strange particle production 
are well described by theoretical models.
However, there are still many details that need improvements,
in particular concerning the treatment of the non-perturbative effects.
The effect of Bose-Einstein correlation has been measured in charged
and neutral kaon final states, and the resulting source radii are found
to be in agreement with values determined in $e^+e^-$ collisions at LEP.
Measurements in the charm sector cover the excited charmed states, 
in agreement with the PDG values, and they have confirmed 
the hypothesis of fragmentation universality.
The searches for pentaquarks in the
doubly strange states at HERA have not turned up any signals,
leading to limits in the production relative to the well known
baryon state $\Xi(1530)^{0}$. 
The prominent structures in the $K^{0}_{S}K^{0}_{S}$ final states
show clear evidence for the $f_0(1710)$ state, consistent
with it being the lowest lying $J^{PC} = 0^{++}$ glueball candidate.

%%%%%%%%%%%%%%%%%%%%%%%%%%%%%%%%%%%%%%%%%%%%%%%%%%%%%%%%%%%%%%%%%%%%%%%%%%
%%%%%%%%%%%%%%%%%%%%%%%%%%%%%%%%%%%%%%%%%%%%%%%%%%%%%%%%%%%%%%%%%%%%%%%%%%

%%%% Here would be an acknowledgement, however the conference organisers
%do not leave enough room in the proceedings - 4 pages for 10 papers.
% somewhere we need to cut - the acknowledgement seems to be most appropriate
% place for this -  we just drop it.
%
%\begin{acknowledgments}
%The authors wish to acknowledge the efforts of the conference organisers.
%Many thanks go to all the colleagues in the H1 and ZEUS collaborations whose
%efforts lead to these many new measurements.
%\end{acknowledgments}

%%%%%%%%%%%%%%%%%%%%%%%%%%%%%%%%%%%%%%%%%%%%%%%%%%%%%%%%%%%%%%%

%\begin{thebibliography}{9}   % Use for  1-9  references

%\tiny

%
\end{document}